\documentstyle{aipproc}

\input BoxedEPS
\SetOzTeXEPSFSpecial
\HideDisplacementBoxes

\newcommand{\etal}{{\em et al.}}

\newcommand{\gevcc}{\hbox{ GeV}\!/\!c^2}
\newcommand{\gev}{\hbox{ GeV}}

\newcommand{\mev}{\hbox{ MeV}}

\newcommand{\kev}{\hbox{ keV}}
\renewcommand{\fm}{\hbox{ fm}}

\def\ltap{\mathop{\raisebox{-.4ex}{\rlap{$\sim$}} 
\raisebox{.4ex}{$<$}}}
\def\gtap{\mathop{\raisebox{-.4ex}{\rlap{$\sim$}} 
\raisebox{.4ex}{$>$}}}

\newcommand{\cfrac}[2]{\textstyle \frac{#1}{#2}}

\def\nogo{\mathop{\not\rightarrow}}

\def\pl#1#2#3{
	{\it Phys. Lett. }{\bf #1}, #2 (19#3)}
\def\prl#1#2#3{
	{\it Phys. Rev. Lett. }{\bf #1}, #2 (19#3)}

\def\prep#1#2#3{
	{\it Phys. Rep. }{\bf #1}, #2 (19#3)}
\def\pr#1#2#3{
	{\it Phys. Rev. D\/}{\bf #1}, #2 (19#3)}

\def\zp#1#2#3{
	{\it Z. Phys. C }{\bf #1}, #2 (19#3)}
\def\ajp#1#2#3{
	{\it Am. J. Phys. }{\bf #1}, #2 (19#3)}
\def\jmp#1#2#3{
	{\it J. Math. Phys. }{\bf #1}, #2 (19#3)}
\def\ap#1#2#3{
	{\it Ann. Phys. (NY) }{\bf #1}, #2 (19#3)}

\def\hepph#1{(electronic archive: hep-ph/#1)}
\newcommand{\hepth}[1]{(electronic archive: hep-th/#1)}

\begin{document}
\title{Realizing the Potential \\ of Quarkonium}

\author{Chris Quigg \thanks{Internet address: \textsf{quigg@fnal.gov}.}}
\address{Fermi National Accelerator Laboratory \thanks{Fermilab is 
operated by Universities Research Association Inc.\ under Contract 
No.\ DE-AC02-76CH03000 with the United States Department of Energy.}
 \\ P.O. Box 500, 
Batavia, Illinois 60510 USA
}

\maketitle

\begin{abstract}
I recall the development of quarkonium quantum mechanics after the 
discovery of $\Upsilon$.  I emphasize the empirical approach to 
determining the force between quarks from the properties of 
$c\bar{c}$ and $b\bar{b}$ bound states.  I review the application of 
scaling laws, semiclassical methods, theorems and near-theorems, and 
inverse-scattering techniques.  I look forward to the next quarkonium 
spectroscopy in the $B_{c}$ system.
\end{abstract}

\section*{Prologue}
I am very happy to share in this celebration of the twentieth 
anniversary of the discovery of the $b$-quark.  The upsilon years were 
a very special time for the development of particle physics.  
Reviewing the events of two decades ago, I was struck not only by the 
pace of discovery, but by how easy it was, and
how much fun we had.  As a young physicist 
of that time, I am grateful not only for the excellent science, but 
also for the excellent people that quarkonium quantum mechanics gave 
me an opportunity to know, work with, and even compete with.  They 
taught me a great deal about physics and life.
\section*{The Empirical Approach}
Charmonium quantum mechanics was already well-launched when the $\Upsilon$ 
came along.  Appelquist \& Politzer \cite{AandP} had shown that 
nonrelativistic quantum mechanics should apply to $Q\bar{Q}$ 
systems.  The Cornell group \cite{kurt} had shown the predictive power 
of the nonrelativistic potential-model approach using a ``culturally 
determined'' potential,
	\begin{equation}
		V(r) = -\frac{\kappa}{r} + \frac{r}{a^{2}}\;\;.
	\end{equation}
Eichten \& Gottfried \cite{EandG} had anticipated the spectroscopy of 
	$b\bar{b}$, predicting
	\begin{eqnarray}
		M(\Upsilon^{\prime}) - M(\Upsilon) & \approx & 420\mev \\
		 & \approx & \cfrac{2}{3} [M(\psi^{\prime}) - M(\psi)]\;\;. \nonumber
	\end{eqnarray}
All this meant that the deductive approach---assuming a form for the 
interquark potential and calculating the consequences---was in very 
good hands.  

Jon Rosner and I were motivated to take a complementary empirical 
approach by the way the quarkonium problem came into our 
consciousness.  The facts we 
	had at our disposal in the summer of 1977 were these:
	 	\begin{center}
		 	\begin{tabular}{|c|c|c|}
				\hline
				E288 & $M(\Upsilon^{\prime}) - M(\Upsilon)$ & $M(\Upsilon^{\prime\prime}) - M(\Upsilon)$  \\
				\hline
				Two-level fit & $650 \pm 30 \mev$ &   \\
				\hline
				Three-level fit & $610 \pm 40 \mev$ & $1000 \pm 120 \mev$  \\
				\hline
				$M(\psi^{\prime}) - M(\psi)$ & 
				$\approx 590 \mev$  & \\
				\hline
			\end{tabular}
	 	\end{center}

We were much impressed with the fact that the 
$\Upsilon^{\prime}-\Upsilon$ spacing is nearly the same	as 
$\psi^{\prime}-\psi$.  In an off-hand conversation with Bernie 
Margolis, who was also visiting Fermilab, Jon asked if he knew what 
kind of potential gave a level spacing independent of the mass of the 
bound particles.  When Bernie said that it was probably some kind of 
power-law, Jon set off on the path that led to a potential $V(r) = 
r^{\epsilon}$ as $\epsilon \rightarrow 0$, and eventually to the 
logarithmic potential.  Accurate numerical calculations and a scaling 
argument showed us that the logarithmic potential, $V(r) = 
C\log(r/r_{0})$ indeed gave a level spacing independent of mass 
\cite{QRlog}.

\begin{figure}[tb]
	\centerline{\BoxedEPSF{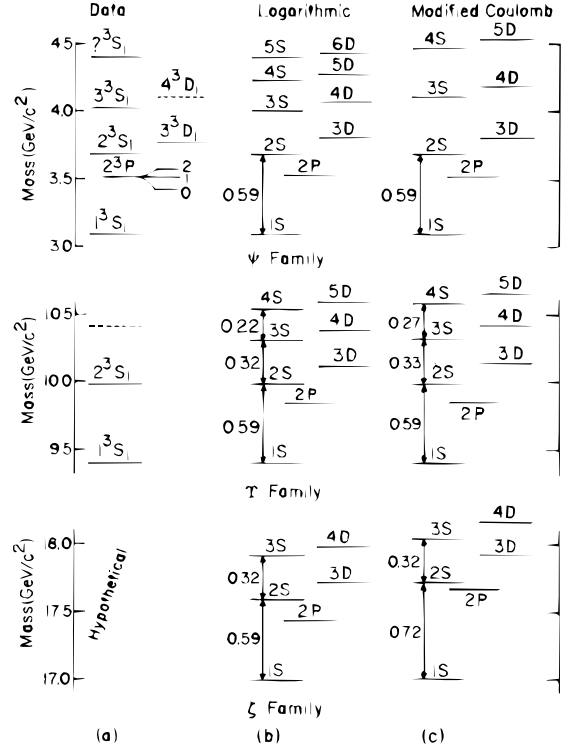  scaled 1000}}
	\medskip
	\caption{Level schemes of the $\psi$, $\Upsilon$, and hypothetical 
	$\zeta$ families in (a) nature, (b) the logarithmic potential, and 
	(c) a $\mathrm{Coulomb}+\mathrm{linear}$ potential.  (From Quigg 
	\& Rosner {\protect \cite{QRlog}}.)}
	{\protect\label{fig:LogLevels}}
\end{figure}

The logarithmic potential gives a good account of the (then known) $\psi$ 
and $\Upsilon$ spectra, as shown in Figure \ref{fig:LogLevels}, but
is not unique in doing so.  It was an easy matter to produce a 
modified Coulomb + linear potential that gave equal spacing for the 
$\psi$ and $\Upsilon$ families, but for no other quark masses.  Now, 
the logarithmic potential is the solution to an idealized statement of 
the experimental facts.  We expect it to be a good representation of 
the interaction in the region of space that governs the properties of 
the narrow $\psi$ and $\Upsilon$ states, but we have no reason to 
attach fundamental significance to it.
It is mildly intriguing that a logarithmic confining interaction emerges from 
the light-front QCD approach when the Hamiltonian is computed to 
second order \cite{lightfront}.

\section*{Scaling the Schr\"{o}dinger Equation}
The Schr\"{o}dinger equation for the reduced radial wavefunction in a 
potential $V(r) = \lambda r^{\nu}$ is
\begin{equation}
	\frac{\hbar^{2}}{2\mu}u^{\prime\prime}+
	\left[E - \lambda r^{\nu} - \frac{\ell(\ell+1)\hbar^{2}}{2\mu 
	r^{2}}\right] u(r) = 0\;\; .
\end{equation}
With the substitutions
\begin{eqnarray}
	r & = & \left(\frac{\hbar^{2}}{2\mu|\lambda|}\right)^{1/(2+\nu)}\rho  
	\;\; ,
	\label{rscale}\\
	E & = & \left(\frac{\hbar^{2}}{2\mu|\lambda|}\right)^{-2/(2+\nu)}
	\left(\frac{\hbar^{2}}{2\mu}\right)\varepsilon \;\; ,\label{escale}
\end{eqnarray}
and the identification $w(\rho) \equiv u(r)$,
we can bring the Schr\"{o}dinger equation to dimensionless form 
\cite{scaling},
\begin{equation}
	w^{\prime\prime}(\rho) +\left[\varepsilon - \mathrm{sgn}(\lambda)\rho^{\nu}
	       -\frac{\ell(\ell+1)}{\rho^{2}}\right]w(\rho) = 0 \;\; .
\end{equation}
The substitution (\ref{rscale}) means that quantities with dimension 
of length scale as $[L] \propto (\mu|\lambda|)^{-1/(2+\nu)}$, 
whereas (\ref{escale}) tells us that quantities with dimensions of 
energy scale as $[\Delta E] \propto 
(\mu)^{-\nu/(2+\nu)}(|\lambda|)^{2/(2+\nu)}$.  The scaling behavior 
in several familiar potentials is shown in Table \ref{table:scale}.
	\begin{table}[tbp]
		\centering
		\caption{How length and energy observables scale with coupling strength 
		$|\lambda|$ and reduced mass $\mu$ in power-law and logarithmic 
		potentials.}
		\begin{tabular}{lll} Potential & $[L]$ & $[E]$ \\
			\hline
			Coulomb & $(\mu|\lambda|)^{-1}$ & $\mu |\lambda|^{2}$ \\
			Log: $V(r)=C\log{r}$ &  $(C\mu)^{-1/2}$ &  $C\mu^{0}$ \\
			Linear & $(\mu|\lambda|)^{-1/3}$ & $\mu^{-1/3}|\lambda|^{2/3}$  \\
			SHO & $(\mu|\lambda|)^{-1/4}$ & $\mu^{-1/2}|\lambda|^{1/2}$  \\
			Square well & $(\mu|\lambda|)^{0}$ & $\mu^{-1}$  \\
			\hline
		\end{tabular}
		\label{table:scale}
	\end{table}

\subsection*{Quarkonium Decays}
The scaling laws have immediate applications to the matrix elements 
that govern quarkonium decays.
Electric and magnetic multipole matrix elements have dimensions
\begin{eqnarray}
	\langle n^{\prime}|{\mathrm{E}}j|n\rangle & \sim & L^{j} , \\
	\langle n^{\prime}|{\mathrm{M}}j|n\rangle & \sim & L^{j}/\mu .
\end{eqnarray}
Radiative widths are given by
\begin{equation}
	\Gamma({\mathrm{E}}j\hbox{ or }{\mathrm{M}}j) \sim p_{\gamma}^{2j+1}
	|\langle n^{\prime}|{\mathrm{E}}j\hbox{ or 
	}{\mathrm{M}}j|n\rangle|^{2}\;\; ,
\end{equation}
so that transition rates scale with mass as 
\begin{eqnarray}
	\Gamma({\mathrm{E}}1) & \sim & \mu^{-(2+3\nu)/(2+\nu)} , \label{e1} \\
	\Gamma({\mathrm{M}}1) & \sim & \mu^{-(4+5\nu)/(2+\nu)} . \label{m1}
\end{eqnarray}
Probability densities have dimensions $L^{-3}$.  Accordingly, the 
wave function squared at the origin scales as 
\begin{equation}
	|\Psi(0)|^{2} \sim \mu^{3/(2+\nu)}\;\; ,
	\label{psilaw}
\end{equation}
so the leptonic width of a vector meson scales as
\begin{eqnarray}
	\Gamma({\mathcal{V}}^{0}\rightarrow \ell^{+}\ell^{-}) & = & 
	16\pi\alpha^{2}e_{Q}^{2}|\Psi(0)|^{2}/M({\mathcal{V}}^{0})^{2} 
	\label{lep}\\
	 & \sim & \mu^{-(1+2\nu)/(2+\nu)}\;\; . \nonumber
\end{eqnarray}
Combining (\ref{e1}) and (\ref{m1}) with (\ref{lep}), we see that
\begin{eqnarray}
	\Gamma(\mathrm{E}1)/\Gamma(\ell^{+}\ell^{-}) & \sim & 
	\mu^{-(1+\nu)/(2+\nu)} , \\
	\Gamma(\mathrm{M}1)/\Gamma(\ell^{+}\ell^{-}) & \sim & 
	\mu^{-3(1+\nu)/(2+\nu)} .  
\end{eqnarray}
For a potential less singular than a Coulomb potential, radiative 
decays become relatively less important than leptonic decays, as $\mu$ 
increases.
\subsection*{Measuring the $b$-quark's charge}
\begin{figure}[tb]
	\centerline{\BoxedEPSF{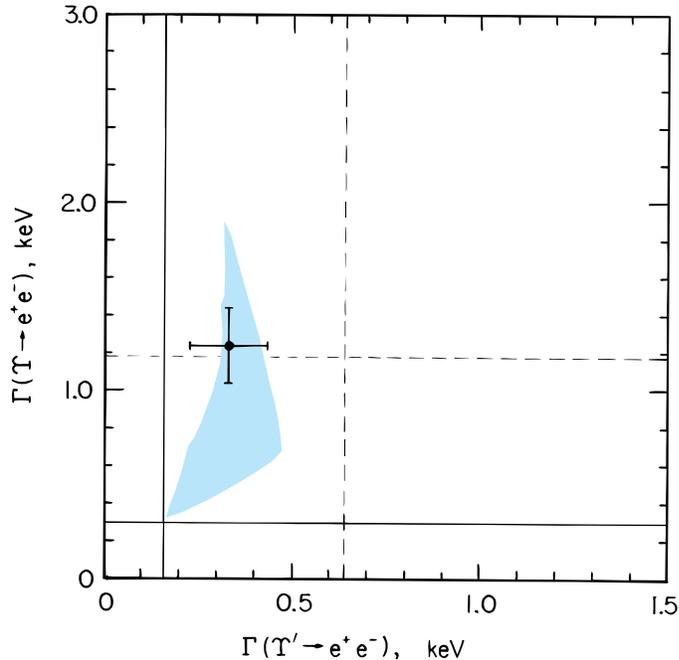  scaled 500}}
	\medskip
	\caption{Expectations for the leptonic widths of $\Upsilon$ and 
	$\Upsilon^{\prime}$.  The lower bounds (\ref{lowerb}) are indicated for 
	$e_{b}=-\frac{1}{3}$ (solid lines) and $e_{b}=\frac{2}{3}$ (dashed 
	lines).  The shaded region shows the widths predicted for 
	$e_{b}=-\frac{1}{3}$ on the basis of twenty potentials from  
	{\protect \cite{inverse2}} 
	that reproduce the $\psi$ and $\psi^{\prime}$ positions and leptonic 
	widths.  The data point represents the 1978 \textsc{doris} results.}
	\protect\label{fig:bcharge}
\end{figure}
For a power-law potential with $\nu \le 1$, the scaling law 
(\ref{psilaw}) implies that
\begin{equation}
	|\Psi_{b}(0)|^{2} \ge \frac{m_{b}}{m_{c}}|\Psi_{c}(0)|^{2}\; ,
\end{equation}
which leads to
\begin{equation}
	\Gamma(\Upsilon_{n} \rightarrow \ell^{+}\ell^{-}) \ge 
	\frac{e_{b}^{2}}{e_{c}^{2}} \cdot \frac{m_{b}}{m_{c}} \cdot 
	\frac{M(\psi_{n})^{2}}{M(\Upsilon_{n})^{2}}\Gamma(\psi_{n} 
	\rightarrow \ell^{+}\ell^{-})\; .
	\label{lowerb}
\end{equation}
The inequality holds for more general potentials than power laws.  We 
can prove it for the ground state for any monotonically increasing potential 
that is concave downward \cite{eb}.  For excited states we have given 
a derivation for the same class of potentials in semiclassical approximation.  

The numerical bounds that follow from (\ref{lowerb}) are indicated in 
Figure \ref{fig:bcharge}.  Measurements presented at the 1978 Tokyo 
Conference by the collaborations working at the \textsc{doris} storage 
ring \cite{doris},
\begin{center}
	\begin{tabular}{ccc}
		$\Gamma(\Upsilon \rightarrow \ell^{+}\ell^{-})$ & = & $1.26 \pm 
		0.21\kev\;\; ,$   \\
		$\Gamma(\Upsilon^{\prime} \rightarrow \ell^{+}\ell^{-})$ & = & $0.36 \pm 
		0.09\kev\;\; ,$   \\
	\end{tabular}
\end{center}
ruled decisively in favor of the $e_{b} = -\cfrac{1}{3}$ assignment.  
The fifth quark was indeed \textit{bottom.}

\subsection*{The Order of Levels}
\begin{figure}[tb]
	\centerline{\BoxedEPSF{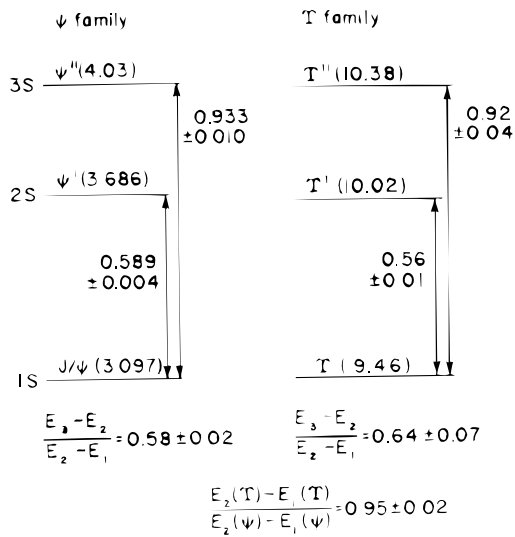  scaled 1000}}
	\smallskip
	\caption{Comparison of $^{3}\mathrm{S}_{1}$ levels of the $\psi$ and 
	$\Upsilon$ families, circa 1979.}
	\protect\label{fig:Martin}
\end{figure}
The discovery in 1975 of the narrow resonances $P_{c}$ and $\chi$ 
confirmed the theoretical expectation that the
	2S and 2P levels are not degenerate in charmonium, as they would be 
	in a pure Coulomb potential.  In response to the question of what 
	the 2S--2P splitting says about the interquark potential, Andr\'{e} 
	Martin and collaborators (R. Bertlmann, H. Grosse, J.-M. 
Richard, and others) constructed a series of elegant theorems on the order of 
levels in potentials \cite{andre,jlrandre}.

After early data on the Upsilons confirmed the similarity of the 
$J\!/\!\psi$ and $\Upsilon$ spectra, as shown in Figure 
\ref{fig:Martin}, Martin used the scaling laws to deduce a simple 
power-law potential, 
\begin{equation}
	V(r) = -8.064\gev + (6.898\gev)(r\cdot1\gev)^{0.1}
	 \;\;. 
\end{equation}
The Martin potential \cite{Martin} has served as a very useful template 
for the $J\!/\!\psi$, $\Upsilon$, and even $\phi(s\bar{s})$ families.  
In addition, it has led to many informative predictions for the masses 
of baryons containing charm and beauty.

\subsection*{A Priority Dispute with Isaac Newton}
About three years ago, Jon Rosner telephoned me to say that Professor 
Chandrasekhar had just advised him that we were in a priority dispute 
with Isaac Newton.  ``Capitulate at once!'' I said.  ``Newton can be 
very terrible.''

In writing his superb commentary on the \textit{Principia Mathematica} 
\cite{chandra}, Chandra had found that Newton was the first to explore 
pairs of dual power-law potentials, and that he had mapped the Kepler 
problem into the harmonic oscillator.  We had shown 
\cite{QRprep}---three centuries later---that the bound-state spectrum 
of an infinitely rising power-law potential, 
\begin{equation} 
	V(r) = \lambda r^{\nu}\;\;(\nu > 0)\;\; ,
\end{equation}
is connected with that of a 
singular potential, 
\begin{equation}
	\bar{V}(r) = \bar{\lambda} 
	r^{\bar{\nu}}\;\;-2 < (\bar{\nu} < 0)\;\;.
\end{equation}
The paired Schr\"{o}dinger equations for the two cases can be written 
as
\begin{eqnarray}
	\frac{\hbar^{2}}{2\mu}u^{\prime\prime}(r) + \left[E - \lambda 
	r^{\nu} - \frac{\ell(\ell+1)\hbar^{2}}{2\mu r^{2}}\right]u(r) & = & 0 
	\;\; , \\
		\frac{\hbar^{2}}{2\mu}v^{\prime\prime}(z) + \left[\bar{E} - \bar{\lambda} 
	z^{\bar{\nu}} - \frac{\bar{\ell}(\bar{\ell}+1)\hbar^{2}}{2\mu 
	z^{2}}\right]v(z) & = & 0 \;\; ,
\end{eqnarray}
where
$(\nu+2)(\bar{\nu}+2) = 4$
and $\bar{E} = \lambda(\bar{\nu}/\nu)^{2}$, $\bar{\lambda} = 
-E(\bar{\nu}/\nu)^{2}$, $(\bar{\ell}+1/2)^{2}\nu^{2} = 
(\ell+1/2)^{2}\bar{\nu}^{2}$, and $z = r^{1+\nu/2}$.

The familiar quantum-mechanical correspondence between the Coulomb  
and harmonic oscillator problems emerges as a special case.  For 
circular orbits, Newton  
cites a relation between $\nu$ and $\bar{\nu}$ equivalent to ours.  
We did capitulate, and so far we have not suffered any indignities at 
Newton's hands.  And because Jon is a scholar, he and Aaron Grant 
have written an excellent historical review of the classical and 
quantum-mechanical analyses \cite{grantr}.

\section*{Counting Narrow Levels of Quarkonium}
\begin{figure}[tb]
	\centerline{\BoxedEPSF{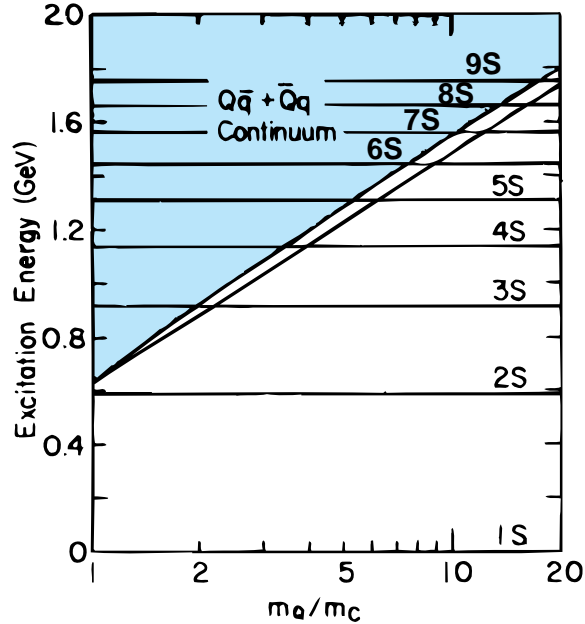  scaled 1600}}
	\caption{Plot of the $Q\bar{q}+\bar{Q}q$ threshold relative to the 
	(ground-state) $1^{3}\mathrm{S}_{1}\;\;Q\bar{Q}$ level as a function of the 
	ratio of the heavy-quark mass $m_{Q}$ to the charmed quark mass 
	$m_{c}$ for a logarithmic potential.  Upper curve: reduced-mass and 
	hyperfine corrections included.  Lower curve (straight line): 
	reduced-mass and hyperfine corrections ignored.  Horizontal lines 
	denote the $2^{3}\mathrm{S}_{1}$, $3^{3}\mathrm{S}_{1}$, \ldots 
	$Q\bar{Q}$ levels in this potential.  (From Quigg and Rosner
	{\protect \cite{counting}}.)}
	\protect\label{fig:Counting}
\end{figure}
Eichten \& Gottfried had argued that a $Q\bar{Q}$ system with 
$m_{Q}\gg m_{c}$ would have at least three narrow 
$^{3}\mathrm{S}_{1}$ levels \cite{EandG}.  As Kurt recalled in his 
talk, this implies a very rich spectroscopy, which figured 
prominently in the scientific case for \textsc{cesr}.  However, the 
observed $\Upsilon^{\prime}-\Upsilon$ spacing is much larger than the 
$420\mev$ they predicted. Figure \ref{fig:Counting} shows that in the
logarithmic potential, we would predict three or four narrow 
$^{3}\mathrm{S}_{1}$ levels of $\Upsilon$, in agreement with Eichten \& Gottfried's 
expectation.  This circumstance led us to ask how general is the 
expectation, and on what does it depend?

Semiclassical methods (whose power Taiji Yamanouchi had impressed on 
us) led us to a remarkable general result 
\cite{counting}: The number 
of narrow $^{3}\mathrm{S}_{1}$ levels is
\begin{equation}
	n \approx 2 \cdot \left(\frac{m_{Q}}{m_{c}}\right)^{1/2}\;\;.
\end{equation}
The derivation is short enough to reproduce in full.

Set the zero of energy at $2m_{Q}$.  The threshold for the 
 dissociation of quarkonium $(Q\bar{Q})\rightarrow Q\bar{q}+\bar{Q}q$ lies 
at an excitation energy $\delta(m_{Q})$.  If $V(r)$ binds $Q\bar{Q}$ 
states rising at least $\delta(m_{Q})$ above $2m_{Q}$, then the WKB 
quantization condition is
\begin{equation}
	\int_{0}^{r_{0}} dr [m_{Q}(\delta(m_{Q})-V(r))]^{1/2} \simeq 
	(n-\cfrac{1}{4})\pi\;\;,
	\label{quant}
\end{equation}
where $r_{0}$ is the point at which $V(r_{0}) = \delta(m_{Q})$.  As 
Eichten \& Gottfried had observed, 
$\delta(m_{Q}) \equiv 2 m({\mathrm{lowest}}\;\;Q\bar{q}\;\;
	{\mathrm{state}})
	 - 2m_{Q}$ approaches a finite limit $\delta_{\infty}$, independent 
	 of $m_{Q}$, as $m_{Q} \rightarrow \infty$.  This means that the only 
	 dependence of (\ref{quant}) on the heavy-quark mass is the explicit 
	 factor of $\sqrt{m_{Q}}$ on the left-hand side.  We have, by 
	 inspection, the general result
	 
	 \begin{equation}
		 (n-\cfrac{1}{4}) \propto \sqrt{m_{Q}}\;\; .
	 	\label{nlev}
	 \end{equation}

\begin{figure}[tb]
	\centerline{\BoxedEPSF{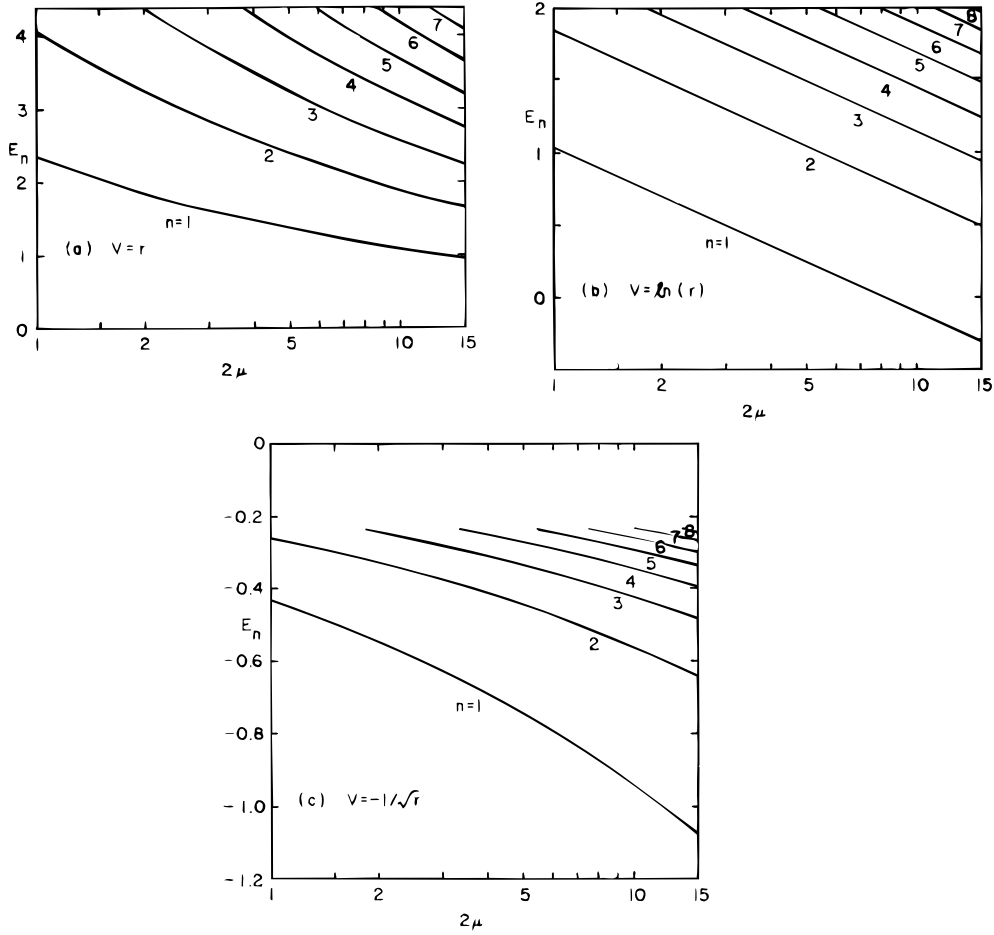  scaled 750}}
	\smallskip
	\caption{Comparison of the reduced-mass dependence of energy levels in 
	three potentials: (a) $V(r) = r$; (b) $V(r) = \ln{r}$; (c) $V(r) = 
	-r^{-1/2}$.}
	\protect\label{fig:Potentials3}
\end{figure}

This universal behavior is realized in different ways for different 
potentials, as illustrated in Figure \ref{fig:Potentials3}.  The 
examples chosen are $V(r) = r$, $V(r) = \ln{r}$, and $V(r) = -r^{-1/2}$,
for which $\Delta E \propto \mu^{-1/3},\; \mu^{0}, \hbox{ and 
}\mu^{+1/3}$, respectively.  All the levels fall deeper into the 
potential as the reduced mass $\mu$ increases, in conformity with 
the Feynman--Hellmann theorem.  For the singular potential with 
$\nu = -\frac{1}{2}$, the levels spread apart as they sink into the 
well.  For the linear potential, the levels are packed more densely 
as $\mu$ rises.  The logarithmic potential is an intermediate case in 
which the level spacing is independent of the mass and all levels 
fall into the well at a common rate given by $E_{i}(\mu^{\prime}) = 
E_{i}(\mu) - \frac{1}{2}\ln{(\mu^{\prime}/\mu)}$.

\section*{Semiclassical Methods and Results}
The power of the WKB approximation for the counting problem 
encouraged us to explore other applications of semiclassical methods.
Evaluating the nonrelativistic connection
\begin{equation}
	|\Psi_{n}(0)|^{2} = \frac{\mu}{2\pi}\left\langle\frac{dV}{dr}\right\rangle_{n}
\label{vir0}
\end{equation}
in the semiclassical approximation, we connect the square of the $s$-wave 
wave function at the origin to the level density:
\begin{equation}
	|\Psi_{n}(0)|^{2} = 
	\frac{(2\mu)^{3/2}}{4\pi^{2}}E_{n}^{1/2}\frac{dE_{n}}{dn}
	\label{psidens}
\end{equation}
(for a nonsingular potential) \cite{wkb}.  For example, in a linear 
potential $V(r) = \lambda r$, $|\Psi_{n}(0)|^{2}$ is independent of $n$, 
by (\ref{vir0}), so $E_{n}\sim n^{2/3}$.

Using the connection
\begin{equation}
	\Gamma_{n} \equiv \Gamma({\mathcal{V}}^{0}_{n} \rightarrow \ell^{+}\ell^{-}) =
	16\pi \alpha^{2} e_{Q}^{2} 
	\frac{|\Psi_{n}(0)|^{2}}{M({\mathcal{V}}^{0}_{n})^{2}} \;\; ,
\end{equation}
we can derive a variety of semiclassical sum rules, including
\begin{equation}
	\sum_{n=\mathrm{narrow}}\frac{\Gamma_{n}}{M_{n}^{p}} \simeq
	\frac{4\alpha^{2} e_{Q}^{2} m_{Q}^{3/2}}{\pi} 
	\int_{0}^{\delta} \frac{dE E^{1/2}}{(2m_{Q}+E)^{2+p}}\;\; ,
\end{equation}
where $\delta = 2M(Q\bar{q})-2m_{Q}$.  These are useful in evaluating 
the heavy-quark mass $m_{Q}$, and in calculating the cross section 
for heavy-quark photoproduction using vector-meson-dominance 
techniques.

The connection (\ref{psidens}) was generalized to higher partial waves 
by Bell and Pasupathy \cite{BandP}, who found

\begin{equation}
	\left|\frac{d^{\ell}R_{n\ell}(r)}{dr^{\ell}}\right|^{2}_{r=0} =
	\frac{1}{\pi} \left[\frac{\ell!}{(2\ell+1)!!}\right]^{2}
	\left(\frac{2\mu E_{n\ell}}{\hbar^{2}}\right)^{\ell+1/2}
	\frac{\partial(2\mu E_{n\ell}/\hbar^{2})}{\partial n} \;\; ,
\end{equation}
and generalized to include singular potentials by Moxhay \& 
Rosner \cite{moxie}.

For a monotonically increasing 
potential, the semiclassical quantization condition
\begin{equation}
	\int_{0}^{r_{0}} dr [2\mu(E-V(r))]^{1/2} = (n-\cfrac{1}{4})\pi
	\label{wkbcond}
\end{equation}
connects the shape of the potential to the level density:
\begin{equation}
	r(V) = \frac{2}{2\mu^{1/2}} \int_{0}^{V} dE (V-E)^{1/2}
	\left[\frac{dE_{n}}{dn}\right]^{-1}\;\; .
	\label{wkbinv}
\end{equation}
Equation (\ref{wkbinv}) is the semiclassical solution to the inverse 
bound-state problem.  Although we never applied it to the quarkonium 
problem, it stimulated us to think of ways to reconstruct the 
interquark potential from the properties of the narrow quarkonium 
levels.

\section*{The Inverse Bound-State Problem}
\begin{figure}[tb]
	\centerline{\BoxedEPSF{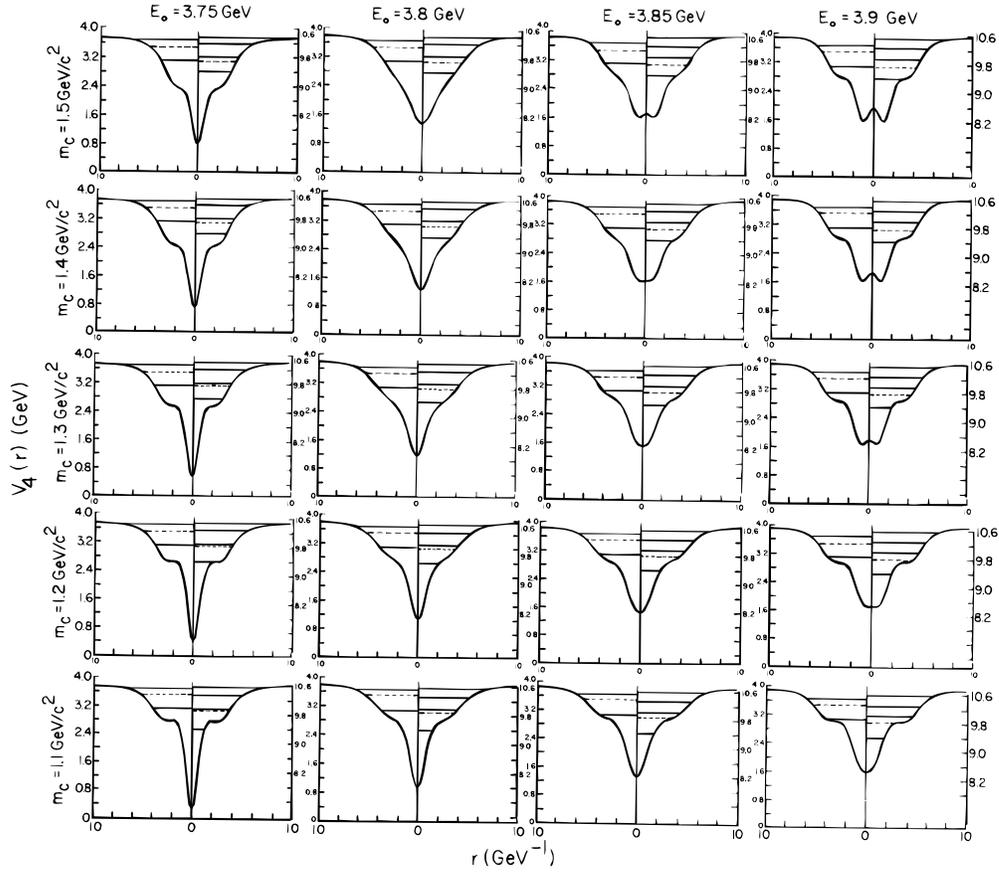  scaled 600}}
	\smallskip
	\caption{Interquark potentials reconstructed from the masses and 
	leptonic widths of $\psi(3.095)$ and $\psi^{\prime}(3.684)$.  The 
	levels of charmonium are indicated on the left-hand side of each 
	graph.  Those of the $\Upsilon$ family are shown on the right-hand 
	side of each graph.  The solid lines denote $^{3}\mathrm{S}_{1}$ levels; 
	dashed lines 
	indicate the $2^{3}\mathrm{P}_{J}$ levels.  (From Thacker, Quigg, and Rosner 
	 {\protect \cite{inverse2}}.)}
	\protect\label{fig:Inverse}
\end{figure}

{ \textit{In one space dimension,} binding energies and phase shifts uniquely 
define a symmetric potential $V(x) = V(-x)$, for which $V(\infty)$ 
approaches a constant (finite) value.  For a ``reflectionless'' 
potential (trivial phase shifts), $V(x)$ is an algebraic function of 
the binding energies \cite{inverse1}.  For the $s$-wave inverse 
problem \textit{in three dimensions,} the central 
potential is implied by binding energies and wave functions at the 
origin.

Thacker, Rosner, and I developed a method of successive approximation 
to confining potentials in terms of a sequence of reflectionless 
potentials that support a finite number of bound states 
\cite{inverse2}.  (It is possible to prove interesting statements about 
the convergence of the procedure \cite{inverse3}.)  Figure 
\ref{fig:Inverse} shows our first attempts to reconstruct potentials 
from what was known about the narrow $c\bar{c}$ levels, and to use 
those potentials to predict the properties of $b\bar{b}$ states.  In 
time, we were able to determine potentials separately from the $\psi$ 
and $\Upsilon$ families, and compare them.  They agree 
remarkably well, except at the shortest distances, to which the 
$\Upsilon$ spectrum has greater sensitivity \cite{inverse4}.

The inverse-scattering approach is free from assumptions about the 
short-distance and long-distance behavior of the potential.  It 
provided additional evidence for flavor independence of the 
interquark potential, and gave us new information on the shape of the 
potential and where it is determined---for $0.1\fm \ltap r \ltap 1\fm$.  
Some important methodological 
improvements have been achieved using techniques of supersymmetric quantum
mechanics \cite{joninv}.

\section*{
Mesons with Beauty and Charm}
\begin{figure}[b!]
	\centerline{\BoxedEPSF{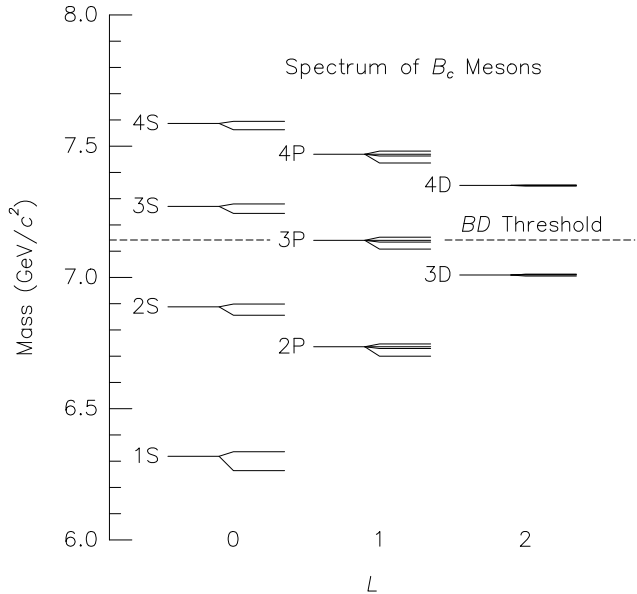  scaled 550}}
	\medskip
	\caption{The spectrum of $b\bar{c}$ states in the Buchm\"{u}ller--Tye 
	potential (after Eichten \& Quigg {\protect \cite{bcspec}}).}
	\protect\label{fig:bcbar}
\end{figure}
The next hurrah for quarkonium physics will be the experimental 
investigation of the $B_{c}$ family of $b\bar{c}$ bound states.  The 
$B_{c}$ family is interesting as a heavy-heavy system that occupies 
the region 
of space between the $J\!/\!\psi$ and the $\Upsilon$.  Since we know 
the heavy-quark potential in that region, we should be able to 
calculate the properties of the $b\bar{c}$ states reliably.  Unlike 
the excited $c\bar{c}$ and $b\bar{b}$ states, all of 
the $b\bar{c}$ levels below $BD$ threshold are stable against strong decays 
to light flavors
($b\bar{c} \nogo \mathrm{gluons}$). They cascade by photonic or 
hadronic transitions to the $B_{c}$ ground state.  The interest in 
these states is not merely academic.  They will soon be discovered and 
studied at the Tevatron Collider through the decays 
$B_{c}\rightarrow \psi\pi, \psi\ell\nu,$ \ldots

Estia Eichten and I have computed the spectrum of 
$b\bar{c}$ bound states in a number of interquark potentials 
\cite{bcspec}.  We find that the mass of the ground state should lie 
in the range $M_{B_{c}} = 6.258 \pm 0.020 \gevcc$.  The low-lying 
levels in the Buchm\"{u}ller--Tye potential \cite{BandT} are shown in 
Figure \ref{fig:bcbar}.  It is noteworthy that approximately 15
$s, p, d$-wave states lie below $BD$ threshold.  \begin{figure}[tb]
	\centerline{\BoxedEPSF{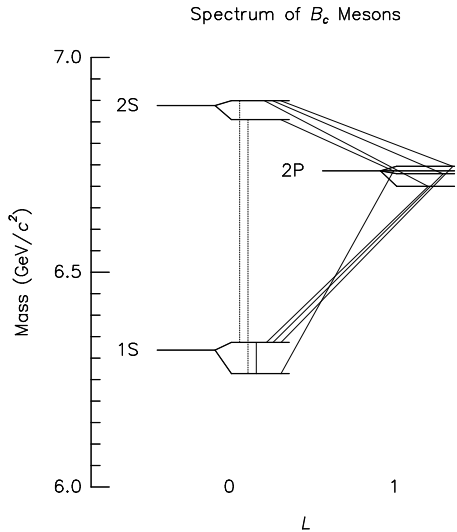  scaled 450}}
	\medskip
	\caption{Prominent transitions in the $B_{c}$ spectrum (after Eichten 
	\& Quigg {\protect \cite{bcspec}}).}
	\protect\label{fig:transitions}
\end{figure}

We have also 
computed the E1, M1, and hadronic transitions between $b\bar{c}$ 
levels.  The transitions involving $n=2$ and $n=1$ states are shown in 
Figure \ref{fig:transitions}.  The narrow widths of these excited 
states are gathered in Table \ref{table:widths}.  We find that the 
deep binding of the heavy quarks has a profound influence on 
decay rates of the $B_{c}$.  For example, we estimate that
$f_{B_{c}} \approx 420\mev \gtap 3f_{\pi}$, which implies that the 
purely leptonic decay $B_{c} \rightarrow \tau \nu_{\tau}$ will be 
unusually prominent.

\begin{table}[b!]
	\centering
	\caption{Total widths of low-lying excited states of $B_{c}$.}
	\begin{tabular}{cc}
			$b\bar{c}$ State & Total Width [keV]  \\
			\hline
			1$^3$S$_1$ & 0.135  \\
			2$^1$S$_0$ & 55  \\
			2$^3$S$_1$ & 90  \\
			2$^3$P$_0$ & 79  \\
			2$(1^+)$ & 100  \\
			2$(1^{+\prime})$ & 56  \\
			2$^3$P$_2$ & 113  \\
			\hline
		\end{tabular}
	\label{table:widths}
\end{table}

In view of the CDF Collaboration's success in reconstructing the 
$\chi_{c}$ and $\chi_{b}$ states, I believe that a 
reasonable---though challenging---experimental goal will be to
map the eight lowest-lying $b\bar{c}$ states (1S, 2S, 2P) through the 
transitions
$2^{3}\mathrm{S}_{1} \rightarrow 1^{3}\mathrm{S}_{1}+\pi\pi$,
$2^{1}\mathrm{S}_{0} \rightarrow 1^{1}\mathrm{S}_{0}+\pi\pi$,
$B_{c}+ 455\hbox{-MeV}\;\gamma$s, and
$(B_{c}^{*}\rightarrow B_{c}\gamma(72\mev)) + 353\hbox{-},
	382\hbox{-}, 397\hbox{-MeV}\;\gamma$s.

Phenomenological issues raised by the $B_{c}$ family include the 
systematics of spin splittings for the unequal-mass $b\bar{c}$ system 
and the importance of relativistic ${\mathcal{O}}(\beta^{2})$ 
corrections.  As a third quarkonium system, $B_{c}$ should provide a 
splendid test of \textit{a priori} calculations from lattice QCD.

\section*{Envoi}
What have we learned from two decades of quarkonium spectroscopy?  The 
first lesson, which underlies all the others, is that nonrelativistic 
quantum mechanics is an appropriate tool for interpreting the quarkonium 
spectra.  Using this tool, we have been able to demonstrate by 
comparing the $c\bar{c}$ and $b\bar{b}$ systems that the force between 
quarks is flavor independent, as we expect from QCD.  Moreover, we 
have been able to map the interaction between heavy quarks in the range 
$0.1\fm \ltap r \ltap 1\fm$.

The potential-model approach allows a predictive spectroscopy, including 
calculations of spin splittings, E1 transition rates, 
the characteristics of $2\mathrm{S} \rightarrow 1\mathrm{S}$ hadronic 
transitions, and the properties of wave functions at the origin, which 
are crucial inputs for calculations of quarkonium production in 
hadronic interactions.

It goes without saying that we have also learned a lot about quantum mechanics!

What can we hope to learn in the years to come?  Among experimental 
goals, we should endeavor to 
complete the charmonium spectrum: refine our knowledge of the $\eta_{c}$ and the 
	$^{3}\mathrm{P}_{J}$ states, confirm the $^{1}\mathrm{P}_{1}$ 
	($h_{c}$) level, find the 
	$\eta_{c}^{\prime}$, and search for narrow D-states.  It is also 
	desirable to expand our knowledge of the $\Upsilon$ spectrum: locate the 
	$\eta_{b}$ and $\eta_{b}^{\prime}$ and the $1^{1}\mathrm{P}_{1}$ 
	($h_{b}$) level, and search for 1D and 2D states.  I am optimistic 
	that we shall soon find the $B_{c}$ and begin to explore the 
	$b\bar{c}$ spectrum.

On the theoretical side, we should be able to refine our understanding 
of relativistic effects and spin splittings.  (Information from the 
$B_{c}$ spectrum should help, in time.)  It may be profitable to 
revisit the coupled-channel effects that influence the spectrum near and 
above the flavor threshold.  There may be lessons of value for 
$B$-factory experiments here.  And finally, we theorists are 
determined to solve QCD and predict the interaction between heavy quarks.

\section*{Acknowledgements}
It is a pleasure to thank Dan Kaplan and our other IIT hosts for the 
invitation to speak and for four pleasant and stimulating days in 
Chicago.  I am grateful to Stew Smith and the Princeton University Physics 
Department for generous hospitality during the spring semester of 1997.  Kyoko 
Kunori provided valuable calligraphic assistance 
[\BoxedEPSF{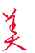  scaled 1000}].  I am indebted to Ken Lane, 
Fermilab Visual Media Services, and the CERN Public Information 
Office for historical photographs shown in my talk.

I am happy to have this opportunity to thank my quarkonium friends 
and collaborators, Hank Thacker, Waikwok Kwong, Jonathan Schonfeld, 
Peter Moxhay, Taiji Yamanouchi, Leon Lederman, Andr\'{e} Martin, and 
especially Jon Rosner and Estia Eichten.  I also want to thank absent 
friends Ben Lee, John Bell, and S. Chandrasekhar for many important 
lessons.

\end{document}